**Superconductivity in $MgHCu_3$ perovskite revisited**

*Piotr Szkudlarek* and Wojciech Grochala**

P. Szkudlarek, Prof. W. Grochala
Center of Novel Technologies, University of Warsaw, Warsaw, Poland
E-mail: p.szkudlarek@cent.uw.edu.pl , w.grochala@cent.uw.edu.pl

***This work commemorates the 115th anniversary of discovery of superconductivity***

Funding: Polish National Science Center (Narodowe Centrum Nauki), grant 2021/41/B/ST5/00195

**Abstract.** We reexamine the crystal structure, electronic structure, lattice dynamics, phonon dispersion, electron-phonon coupling, and superconducting properties of $MgHCu_3$ perovskite using the PBEsol and PBE functionals. This perovskite phase was recently proposed to exhibit superconductivity with the critical superconducting temperature, $T_C$, of 42 K, which falls slightly over the classical 40 K limit for the phonon-driven superconductivity. We show that although the crystal and electronic structure of this hypothetical compound are quite robust with respect to the k-point mesh and functional used, yet the phonons and phonon-related properties are extremely sensitive to the density of the grid chosen as well as functional used for calculations. Correspondingly, the values of the critical superconducting temperature calculated here for different Gaussian broadenings vary in a broad range of ca. 10–31 K and they do not exceed the classical limit. We suggest that the properties of this and many other high-$T_C$ hydrides claimed should be thoroughly scrutinized using a variety of functionals, and benchmarked with experiment, to provide more reliable values of $T_C$.

## 1. Introduction

The quest for the high-temperature (high-$T_C$) superconductivity (SC) has long been focused on elemental hydrogen[1] and hydrogen-rich materials[2,3,4,5]. This is due to the fact that hydrogen is the lightest chemical element and thus it provides the largest preexponential factor (Debye frequency) in the expression for the superconducting critical temperature ($T_C$) value within the BCS theory[6,7]. Indeed, after many decades of intense quest, high-$T_C$ SC has been found in compressed $H_3S$[8], $LaH_{x\sim10}$ [9,10] and similar polyhydrides[11,12] at very high external pressures. The largest reliably confirmed $T_C$ value reported so far is 250-260 K for $LaH_{x\sim10}$ and $YH_{x\sim9}$ at pressures exceeding 150 GPa (1.5 mln atm) and there seem to be some room for further increasing this value up to 330 K[13]. While these findings are of immense scientific value *per se*, they cannot be immediately transferred to any practical applications due to the pressure range required for stability of these phases. The diamond anvil cells now provide the only technical mean to achieve such immense compression and the sample size is limited to few dozens μm at best. Simultaneously, the SC disappears if samples are quenched below a certain pressure, mostly due to decomposition of the polyhydride with the simultaneous release of $H_2$.

On the other hand, the SC reported so far in hydride materials at ambient pressure does not exceed the 20 K threshold. The experimentally confirmed examples encompass $Th_4H_{15}$[14] ($T_C$=8.3 K), $PdH_x$[15] ($T_C$= 9 K), $ZrH_3$[16] ($T_C$= 11.6 K) and $(Pd,Ag)H_x$[17] ($T_C$= 16 K). This is obviously disappointing and therefore – given a rich chemistry of metal hydrides[18] – many theoretical studies have been recently devoted to novel stoichiometries which could deliver higher $T_C$ values. Thus, SC was predicted to occur, inter alia, for a single layer of the hole-doped LiH[19] ($T_C$=17.5 K), $PdCuH_2$[20] ($T_C$=34 K), $MgHCu_3$[21] ($T_C$=42 K), $KInH_3$[22] ($T_C$=72.9 K), $SrBe_4H_{12}$[23] ($T_C$= 76.7 K), $CsBH_5$[24] ($T_C$=83 K), $Li_2CuH_6$[25] ($T_C$=86 K), $(Mg,Na)(BH_4)_2$[26] ($T_C$=100 K), $Li_2AgH_6$[27] ($T_C$=108.8 K), $PdH_4$[28] ($T_C$=133–146 K), and $Mg_2IrH_6$[29] ($T_C$=160 K), to mention just a few systems[30]. The authors of the recent work conclude that systems such as $Li_2AgH_6$ "*likely approach the practical limit for conventional superconductivity at ambient pressure*"[27], however, the value obtained for $Mg_2IrH_6$ cannot be neglected. This suggestion, if true, would mean that it is possible to **substantially** (i.e. 4-fold) surpass the classical 40 K limit which was long believed to constitute an upper limit of BCS-type materials at ambient pressure. The quoted statement is rather surprising given the fact that the vast majority of the proposed stoichiometries are vastly thermodynamically unstable; note that the preparation of metastable hydrides at ambient pressure is immensely difficult because H is the only atom which can effectively move through the crystal lattice via (in essence temperature-independent) quantum tunnelling. Notable exceptions from this rule of thumb are provided by, *inter alia*, semiconducting CuH, $ZnH_2$ and $FeH_x$, which can be prepared in the laboratory although they are susceptible to decomposition via thermal or mechanical initiation[31,32,33]. The occurence of metastable semiconducting or insulating hydrides – already very rare – is even more rare among the metallic hydrides (only some of those may be prepared at elevated pressure and quenched to ambient pressure conditions).

Here, we take a closer look at $MgHCu_3$[21] with its simple perovskite structure and the previously computed $T_C$ value of 42 K, which is only slightly larger than the above-mentioned classical limit, but still quite impressive given that it is a relatively H-poor hydride. We reexamine its key parameters related to the SC behaviour and we compare them with those given in the original report[21].

## 2. The key parameters of SC behaviour

Our studies employed the Generalized Gradient Approximation and two types of functional, PBEsol[34] (used by the authors of the original work, and for comparison a related PBE[36], and as well as several different k-point meshes, both less and more dense than the originally used one (15x15x15). The k-mesh reported in the manuscript refers to the density of mesh used for the SCF calculations. For every system, the q-mesh density (onto which phonons are computed via DFPT) can be obtained by dividing the number density of the scf-mesh by 3, while the density of the denser k-mesh (for electron-phonon double-delta integration) can be obtained by multiplying the density of the original mesh by 4. For PBEsol calculations we have employed norm conserving scalar relativistic Vanderbilt pseudopotentials ver.4 from https://www.pseudo-dojo.org/, while pseudopotentials for PBE calculations were supplied to us by one of the authors[35]. We always report the values of $T_C$ for the Gaussian broadening of 0.03 Ry[19], which yielded reasonable convergence of the $T_C$ value; however, we show the $T_C$ – broadening dependence in the electronic supplement. The $T_C$ value was computed using the McMillan–Allen–Dynes formula,[37,38] derived from the Eliashberg theory.[39] We have always adopted here $\mu^* = 0.1$.

In **Table 1** we list the key parameters of the $MgHCu_3$ superconductor, i.e. the calculated cubic lattice constant, ***a***, the electronic density of states at the Fermi level, ***DOS($E_F$)***, the logarithmic average frequency, $\boldsymbol{\varpi_{log}}$, the electron-phonon coupling constant, $\boldsymbol{\Lambda}$, and the resulting $\boldsymbol{T_C}$ value.

**Table 1.** The key parameters of the $MgHCu_3$ superconductor, i.e. the calculated cubic lattice constant, ***a***, the electronic density of states at the Fermi level, ***DOS(E_F)***, the logarithmic average frequency, $\varpi_{log}$, the electron-phonon coupling constant, ***Λ***, and the resulting $T_C$ value.

| k-point mash | F[a)] | a [Å] | $DOS(E_F)$ [states (eV unit cell)$^{-1}$] | $\varpi_{log}$ [K] | Λ [1] | $T_C$ [K] | Ref. |
|---|---|---|---|---|---|---|---|
| 15x15x15 | PBEsol | 3.77 | 1.2 | 402 | 0.83 | 42 | [21] |
| 12x12x12 | PBEsol | 3.78 | 1.17 | 384 | 0.75 | 15.5 | This work |
| 15x15x15 | PBEsol | 3.78 | 1.18 | 374 | 0.86 | 20.1 | This work |
| 18x18x18 | PBEsol | 3.78 | 1.17 | 365 | 0.85 | 19.6 | This work |
| 24x24x24 | PBEsol | 3.78 | ND[b)] | 423 | 0.60 | 9.5 | This work |
| 12x12x12 | PBE | 3.84 | 1.33 | 291 | 1.42 | 31.3 | This work |
| 15x15x15 | PBE | 3.84 | 1.23 | Structure dynamically unstable | | | This work |
| 18x18x18 | PBE | 3.84 | 1.26 | Structure dynamically unstable | | | This work |
| 24x24x24 | PBE | 3.84 | 1.20 | 158 | 2.95 | 28.3 | This work |

[a)] F – functional; [b)] ND – not determined.

### 2.1. Lattice constant, electronic DOS and DOS($E_F$).

It can be seen from Table 1 that the lattice constant optimized here using the PBEsol functional (3.78 Å) is nearly identical to that reported by these authors (3.77 Å)[21]. We also notice that it is nearly independent on the k-point mesh in our study. However, a slightly (i.e.; 1.6%) larger value of 3.84 Å larger is obtained with the PBE functional (and also quite insensitive to the k-point grid density). Such differences fall perfectly within the errors of the DFT methods.

Importantly, the electronic structure calculated here is immensely similar to that reported before. In addition, the electronic band structure is only weakly dependent on the k-point mash density and the functional chosen (Figure 1, *cf.* S1 in SI).

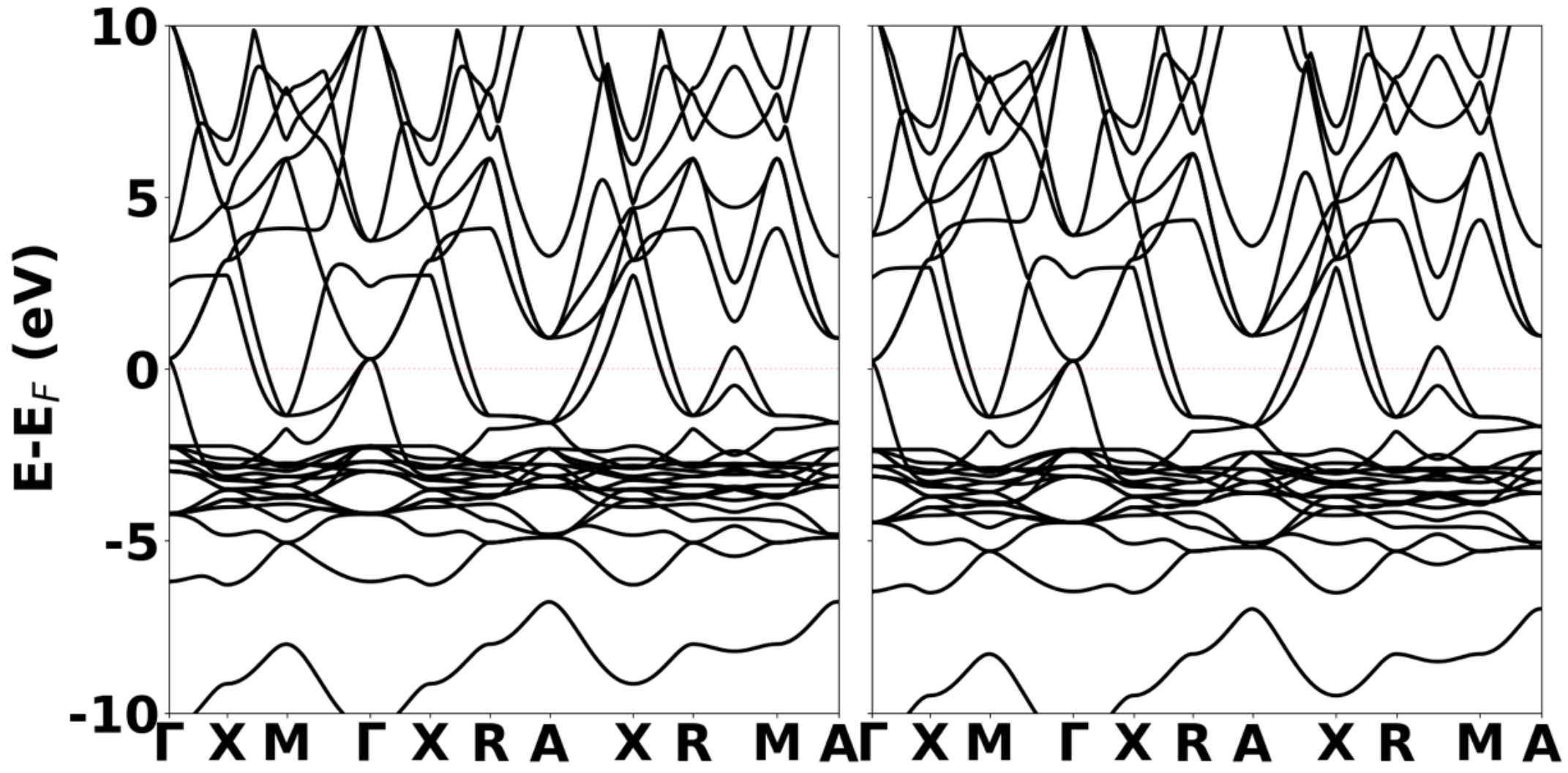

**Figure 1.** The electronic band structure for the densest studied k-point mesh (24x24x24) obtained using the PBE (left) and PBEsol functional (right).

Consequently, also the values of DOS at the Fermi level do not differ much for diverse k-meshes and functionals. In general, these values are slightly larger for the PBE functional (1.20–1.33 states $eV^{-1}$ unit $cell^{-1}$) than for the PBEsol (1.17–1.20 states $eV^{-1}$ unit $cell^{-1}$), which is expected since the lattice constant for the former is slightly larger than for the latter. All values are close to that reported previously (DOS at the Fermi level of 1.20 states $eV^{-1}$ unit $cell^{-1}$). For PBE (24x24x24) the contribution of H to the DOS at the Fermi level is *ca.* 6.7% (when dividing H partial DOS by total DOS) or *ca.* 10.7% (when dividing H partial DOS by the sum of all partial DOS values). The latter is closer to the value of 9.5% reported in[21], which suggests that the authors[21] have probably applied the latter technical formula.

**2.2. Phonon spectra, dynamic stability and Eliashberg function.**

To our surprise, we found out that the calculated phonon spectra are extremely sensitive to the choice of the functional and especially the k-point grid (*cf.* S2 in SI). In Figure 2 we show the phonon dispersion for the PBE functional and all four k-point grids studied. In Figure 3 we present the comparison of the phonon dispersion and phonon density of states for the densest studied k-point mesh (24x24x24) obtained using the PBEsol and PBE functionals.

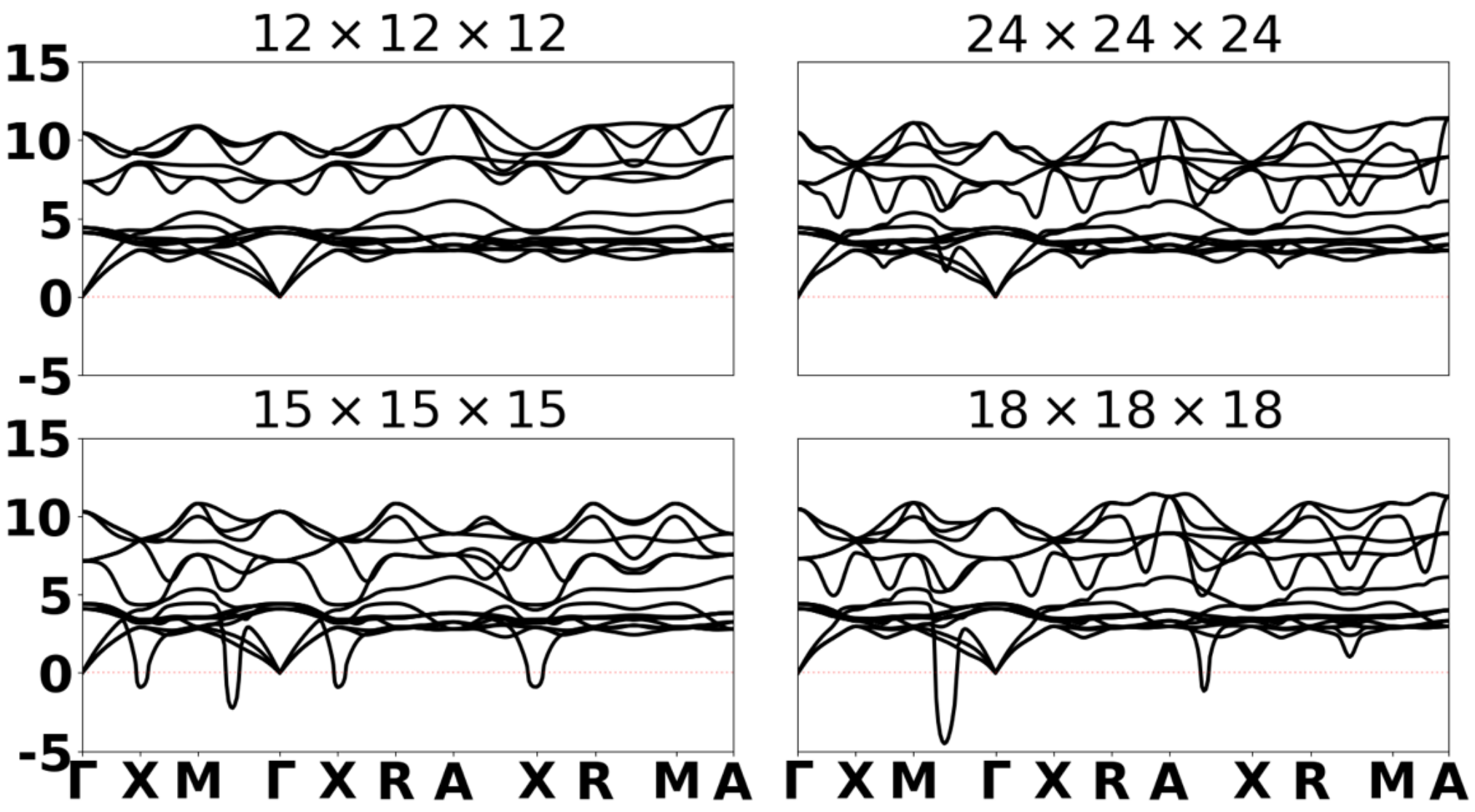

**Figure 2.** The phonon dispersion curves obtained using the PBE functional and the four different grid densities (increasing counterclockwise).

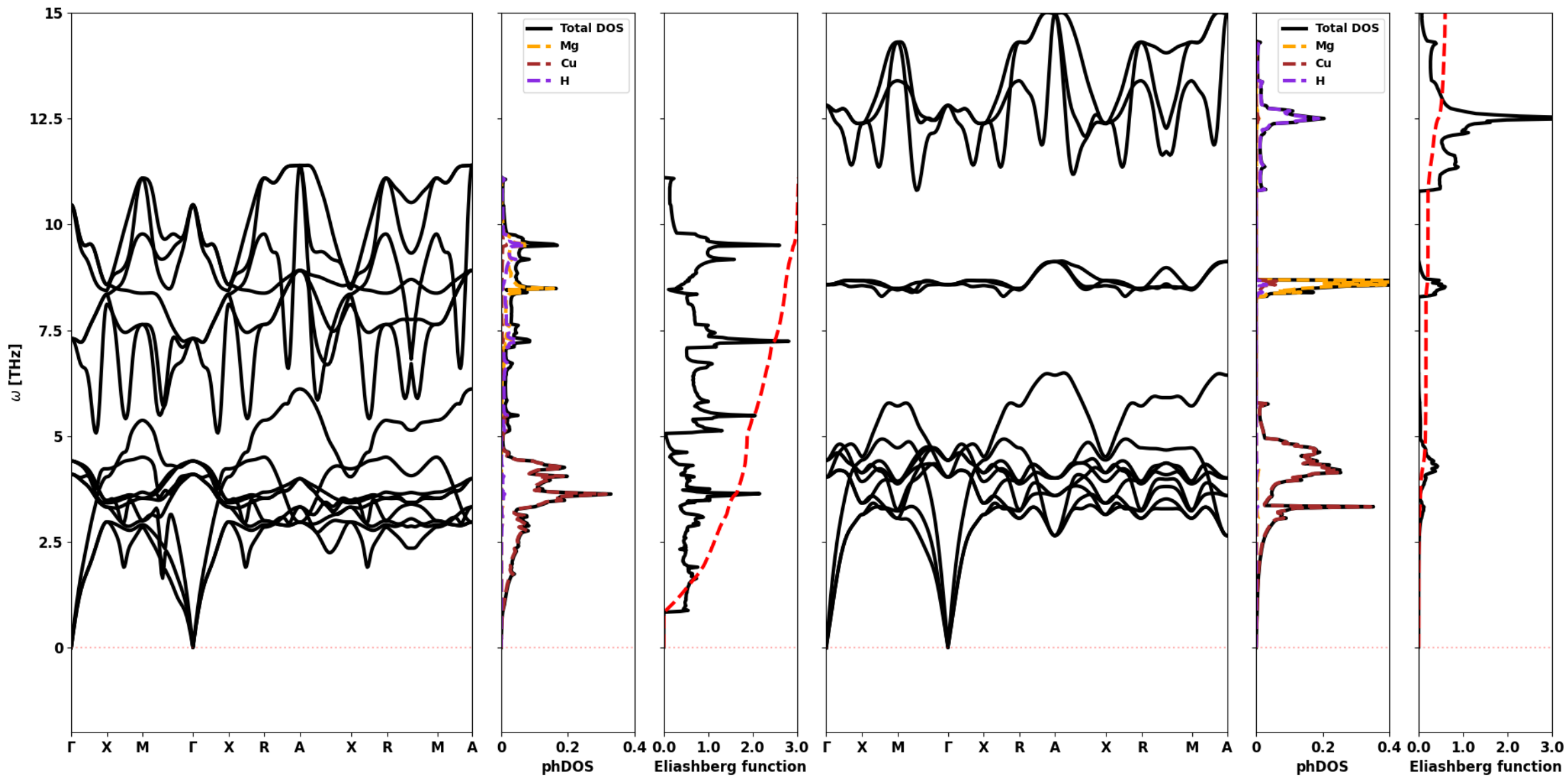

**Figure 3.** The phonon dispersion curves and DOS, as well as the Eliashberg function for the densest studied k-point mesh (24x24x24) obtained using the PBE (left) and PBEsol functionals (right).

The phonon spectrum computed for PBE functional and the 12x12x12 k-point mesh reveals no dynamic instabilities. The phonon branches may be grouped roughly in three frequency regions with the acoustic and lowest frequency optical phonons from 0 to 6.1 THz. The higher-frequency modes are found in the 6.1-9 THz and 9-12.2 THz windows. Importantly, when the 15x15x15 mesh is applied, the spectrum changes considerably. First, the two upper branches broaden and the maximum frequency drops to 10.8 THz. Secondly, imaginary modes appear at X(0,0,½), and in between M(½,½,0) and Gamma (0,0,0). The largest imaginary frequency is ca. *i*2.3 THz. When the 18x18x18 mesh is used, the phonon spectrum changes further, with the most pronounced imaginary frequency in between M and Gamma increasing to *i*4.5 THz. Finally, when the densest 24x24x24 mesh is used, the phonon dispersion regains all real frequencies, but its shape is quite different in details from the 12x12x12 one. One important difference is that for the densest mesh all phonon branches are broadened, which cases the phonon density of states to have non-zero values for the entire spectral range from 0 to 11.4 THz. Also, the uppermost and middle frequency regions interpenetrate much more with one another than for the 12x12x12 mesh.

The fact that calculations predict lack of dynamic stability of some intermediate mesh, and stability for the least dense and also for the densest one ("re-entrant stability") may point out to the vicinity of a phonon-driven phase transition for $MgHCu_3$. Obviously, the presence of soft phonons (close to the phase transition[40]) may enhance SC behaviour but if instead of being soft a phonon become imaginary then this most often leads to either substantial decrease of the $T_C$ value or even to a band gap opening (i.e., $T_C$ = 0 K[41,42]) for the distorted lower-energy and lower-symmetry structure. The borderline between high-$T_C$ superconductivity and the metal-insulator transition is indeed very narrow, as testified by many systems, and proper theoretical description of this thin boundary is challenging.

Since generally for a proper description of metals one should use as dense k-point mesh as possible, we compared the results obtained using the densest grid but two different functionals. We notice from Figure 3 that the phonon dispersion curves computed with the PBEsol and PBE functionals are quite different from each other. First, the uppermost frequency for PBEsol is close to 15.0 THz, while that for PBE is only 11.4 THz. Secondly, the intermediate-frequency branches which for PBE form a low-DOS broad-band net between 6.1 and 9.0 THz, are now extremely localized for PBEsol, with very narrow bands and large phonon DOS around 8.6 THz (mostly coming from motions of Mg atoms). Thus, qualitatively, two related functionals give dramatically different description of the lattice dynamics. We note, however, that our results for the PBEsol functional resemble those reported earlier with the same functional[21].

**2.3. Logarithmic average frequency, electron-phonon coupling constant and the resulting $T_C$ value.**

Since the phonon spectra turned out to be crucially dependent on the functional chosen and the k-mesh, it is not surprising that also the logarithmic average frequency and electron-phonon coupling constant are heavily dependent on these factors. Thus, the values of $\varpi_{log}$ vary between 158 K (PBE, 24x24x24 mesh) and 423 K (PBEsol, 24x24x24 mesh) (these authors derived the value of 402 K[21]). On the other hand, the electron-phonon coupling constant, $\Lambda$, varies here between 0.75 (PBEsol, 12x12x12 mesh) and 2.95 (PBE, 24x24x24 mesh). These are immense differences as they fall into "weak" and "very strong" coupling regimes, respectively.

Since the $T_C$ value is crucially dependent on all the previously discussed parameters, it is not surprising at all that its value varies significantly depending on the k-point mesh as well as the functional used. Specifically, the lowest $T_C$ value of 9.5 K was calculated here for PBEsol functional and the densest grid, while the largest one of 28.3 K for PBE functional, and also the densest k-point mesh (all values are reported for the Gaussian broadening of 0.03 Ry; for full dependence *cf.* S3 in SI). We also notice that since in the latter case $\Lambda > 1.5$, the Migdal-Eliashberg equations do not work well[43,44], so the precise $T_C$ value in fact remains unknown. None of our values approaches the 42 K one given in [21] for PBEsol and the 15x15x15 mesh, and both remain within the "classical limit". Recalling that proper description of metals requires as dense mesh as possible one could naturally be prone to trust the results for the 24x24x24 mesh more than for less dense ones, but huge difference between the $T_C$ values computed with two different functionals completely annihilates such trust. While some arguments could be said in favour of the PBEsol functional ("revised for solids") with respect to the more classical PBE, yet obviously the construction of these functionals was never benchmarked against the $T_C$ values but rather against less sensitive parameters such as lattice constants etc.

## 3. Conclusions

The large breadth of the $T_C$ values computed for $MgHCu_3$ perovskite with different functionals and grids suggests that these calculations desperately need experimental benchmarking. Since DFT is in practice a semiempirical method, extremely thorough theoretical scrutiny of superconducting properties is also needed, especially utilizing functionals which can handle well the propensity towards localization/delocalization of the electronic density, such as e.g. HSE06. In addition, anharmonicity should be taken into account as it may strongly impact the results in some cases. Until then, as Neil Ashcroft used to joke, "*all theories of superconductivity have one parameter only, and that parameter is the $T_C$ value*"[45]. Only when experiment successfully confirms the 42 K value[21] we will probably be able to formulate claim whether the classical 40 K limit could indeed be surpassed in metal hydrides.

## Methods

*Computational Details*: Calculations were performed with the Quantum Espresso suite[46,47]. Derivation of the $T_C$ values employed matdyn.x as in[19].

## Acknowledgements

W.G. is grateful to the Polish National Science Center for project OPUS Chem-Cap (2021/41/B/ST5/00195). Computations were performed at the Interdisciplinary Center for Mathematical and Computational Modeling of Warsaw University (grant SAPPHIRE GA83-34) and at the Wrocław Centre for Networking and Supercomputing (grant no. 484).

**Supporting Information**

Supporting Information is available from the Wiley Online Library or from the author. It contains electronic dispersion and density of states and phonon dispersion and density of states for all k-point meshes and functionals. Also, for dynamically stable structures we present the Eliashberg function overlapped with the phonon DOS as well as the dependence of the calculated $T_C$ on the Gaussian broadening.

# Supporting Information

## Superconductivity in $MgHCu_3$ perovskite revisited

*Piotr Szkudlarek* and Wojciech Grochala**

S1. Electronic band structure and DOS computed with PBE or PBEsol functionals and four different increasing grid densities.

PBE 12x12x12

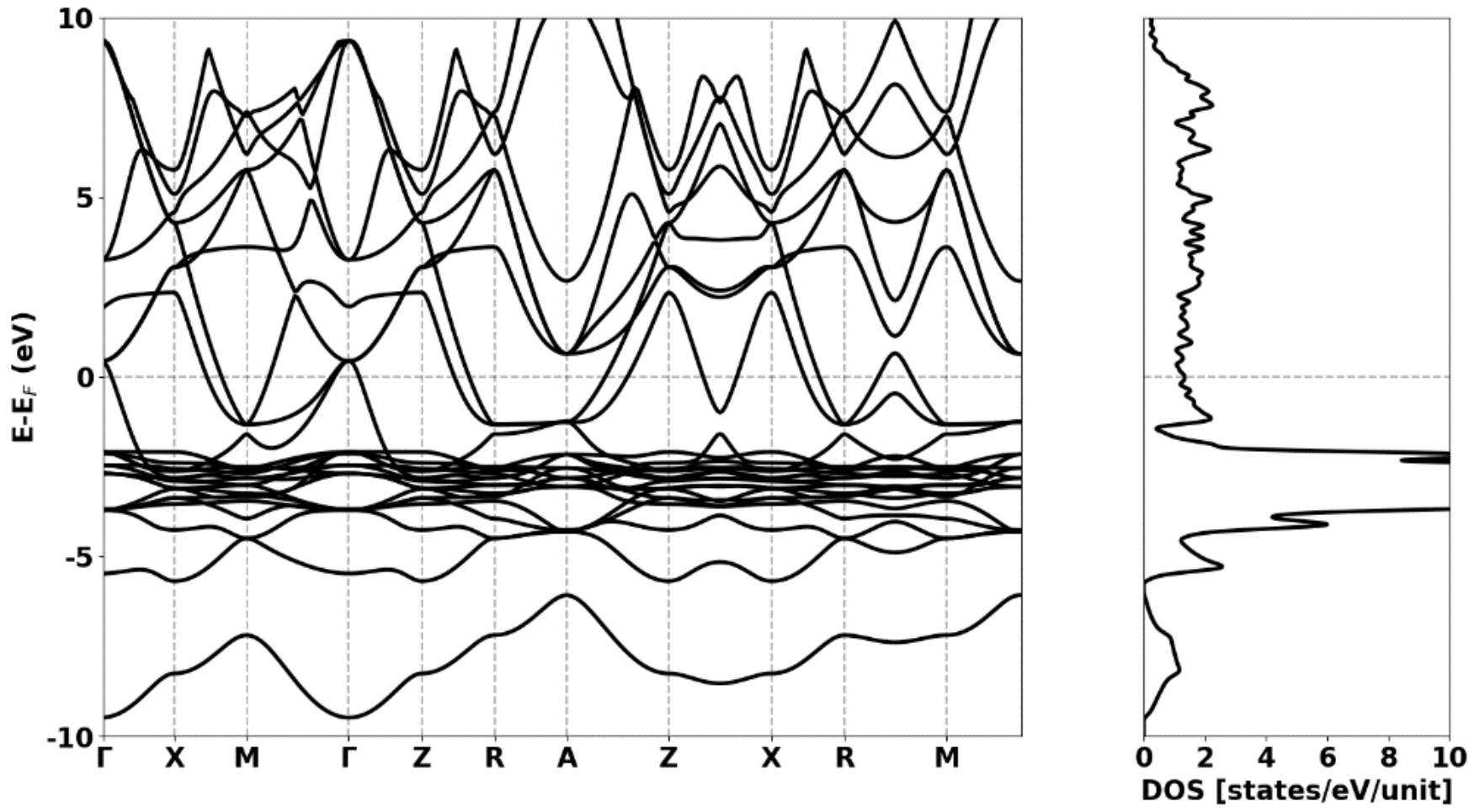

PBE 15x15x15

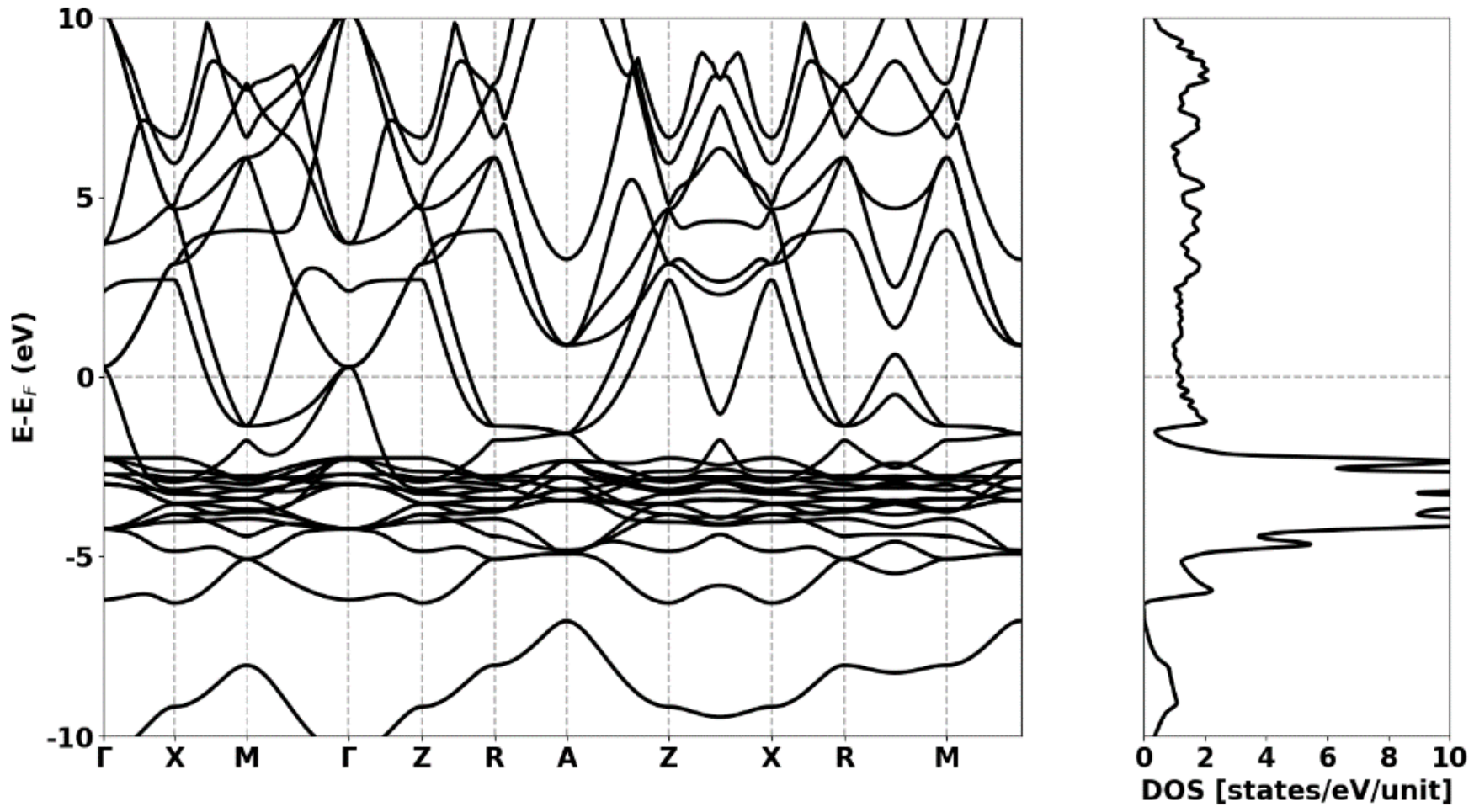

PBE 18x18x18

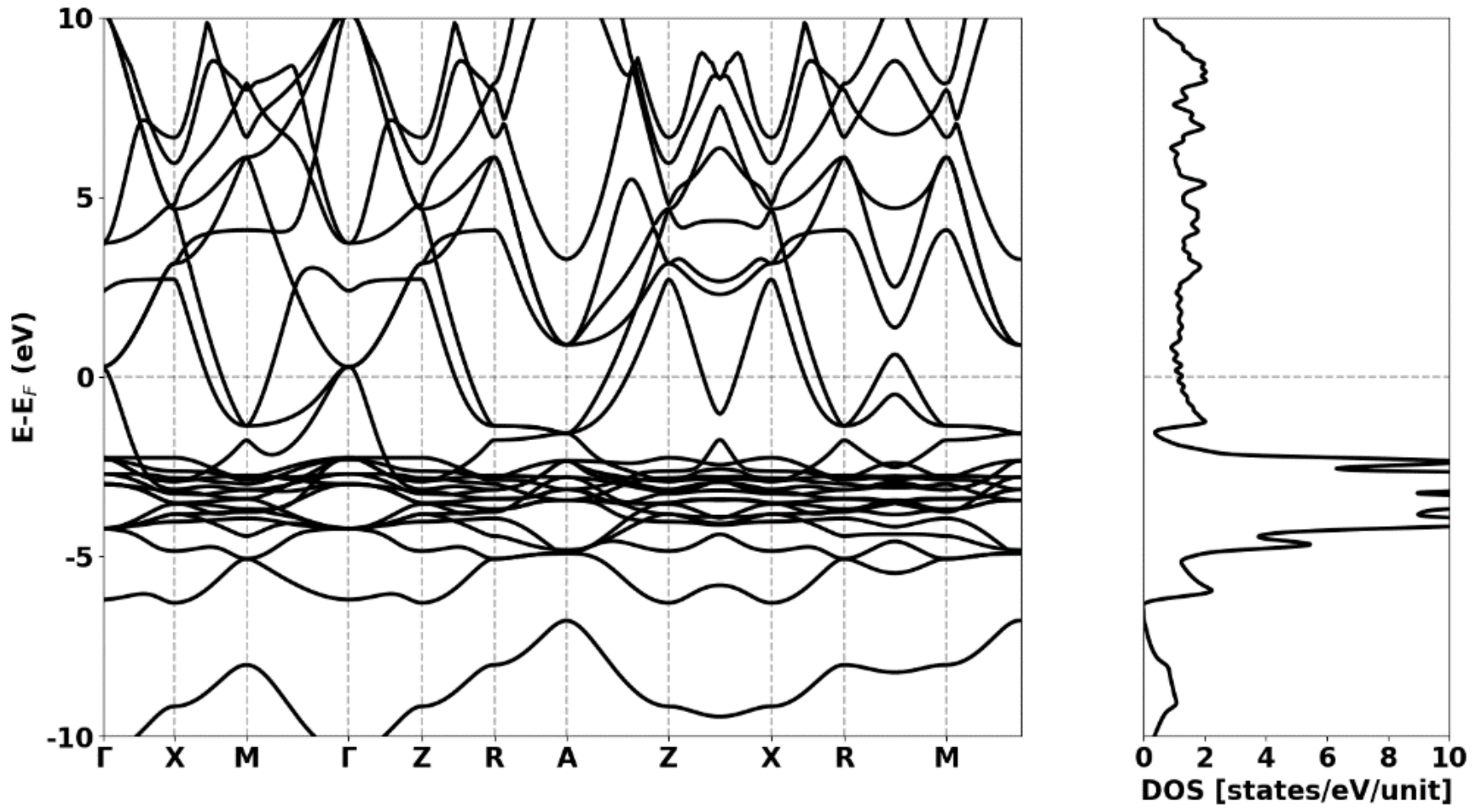

PBE 24x24x24

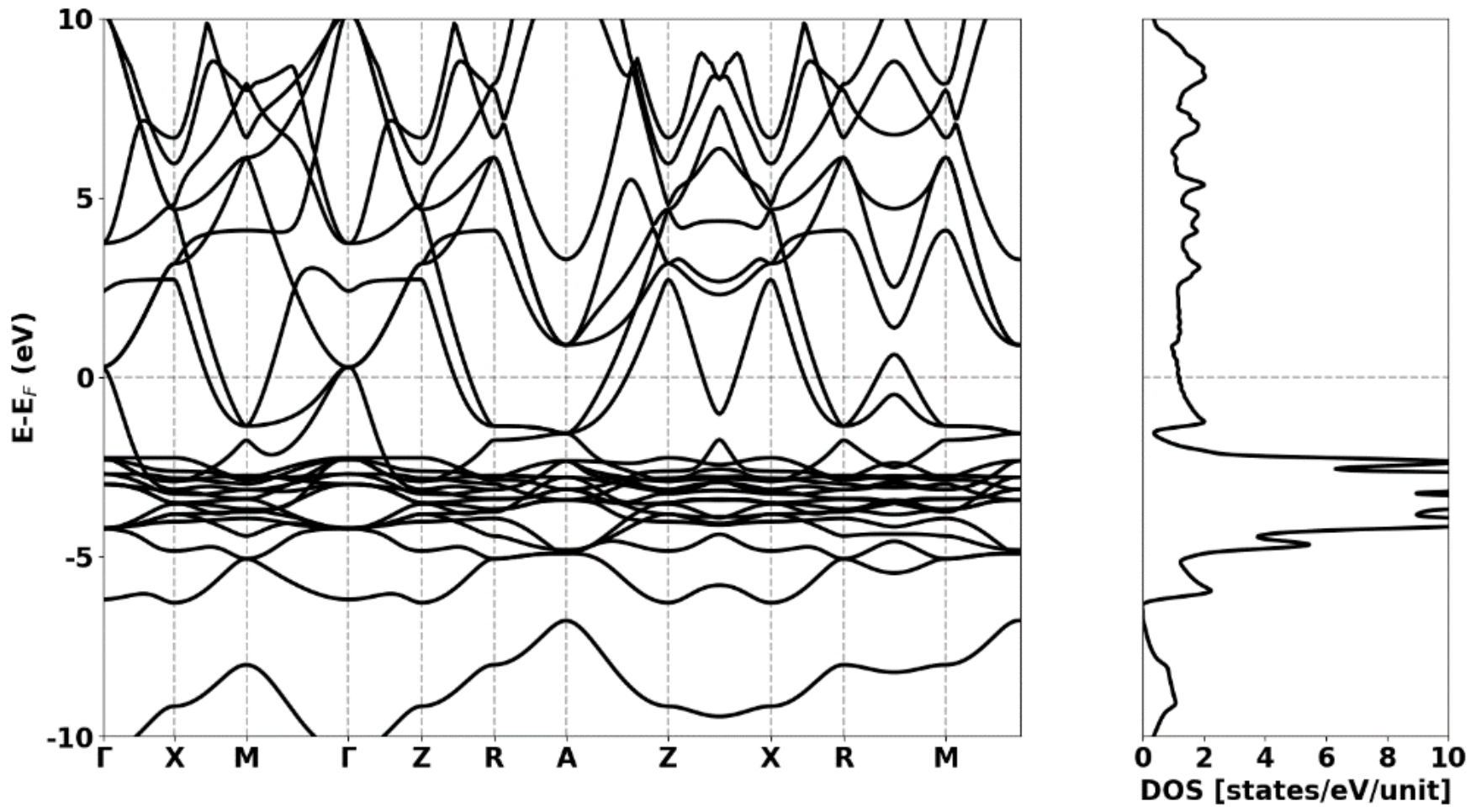

PBEsol 12x12x12

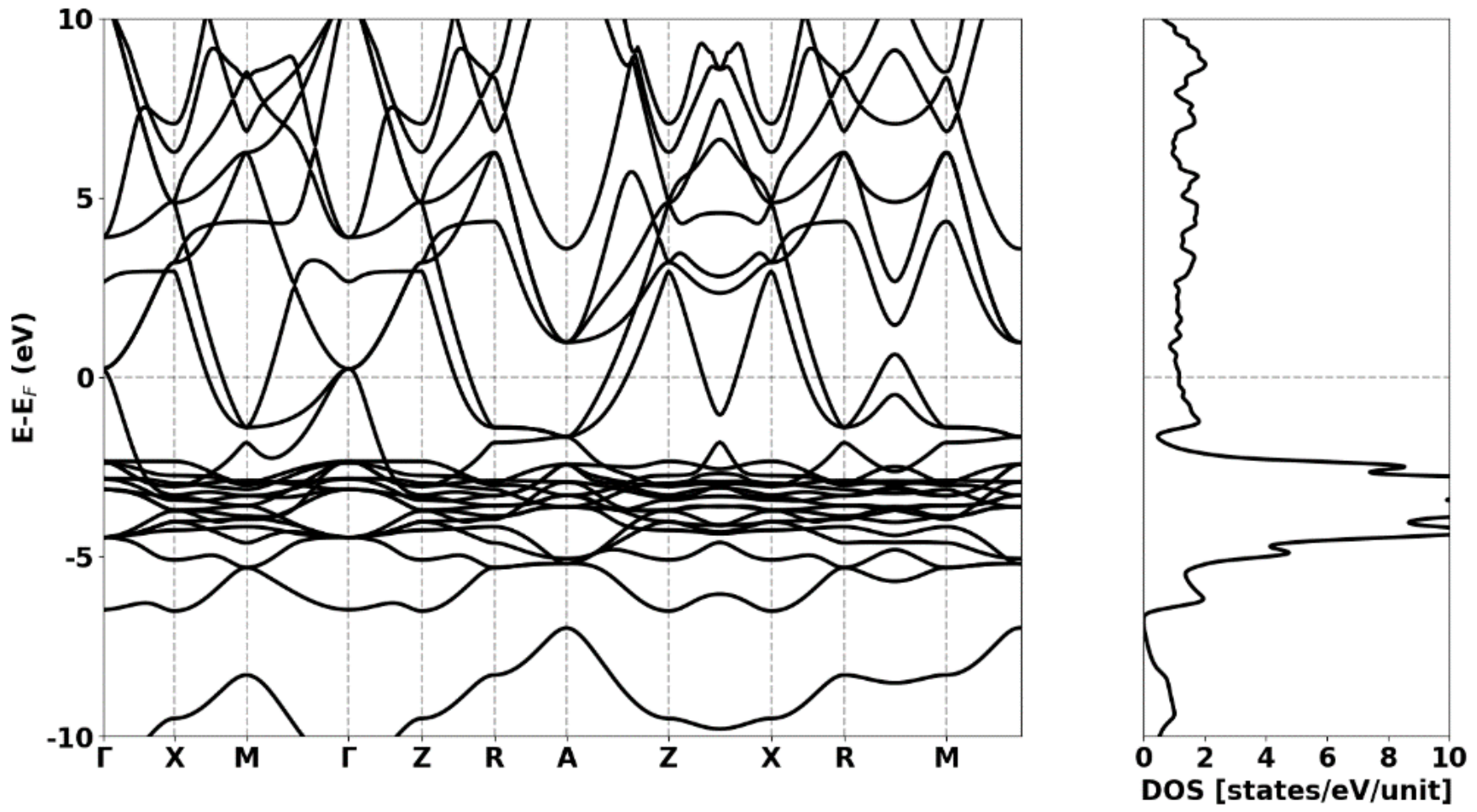

PBEsol 15x15x15

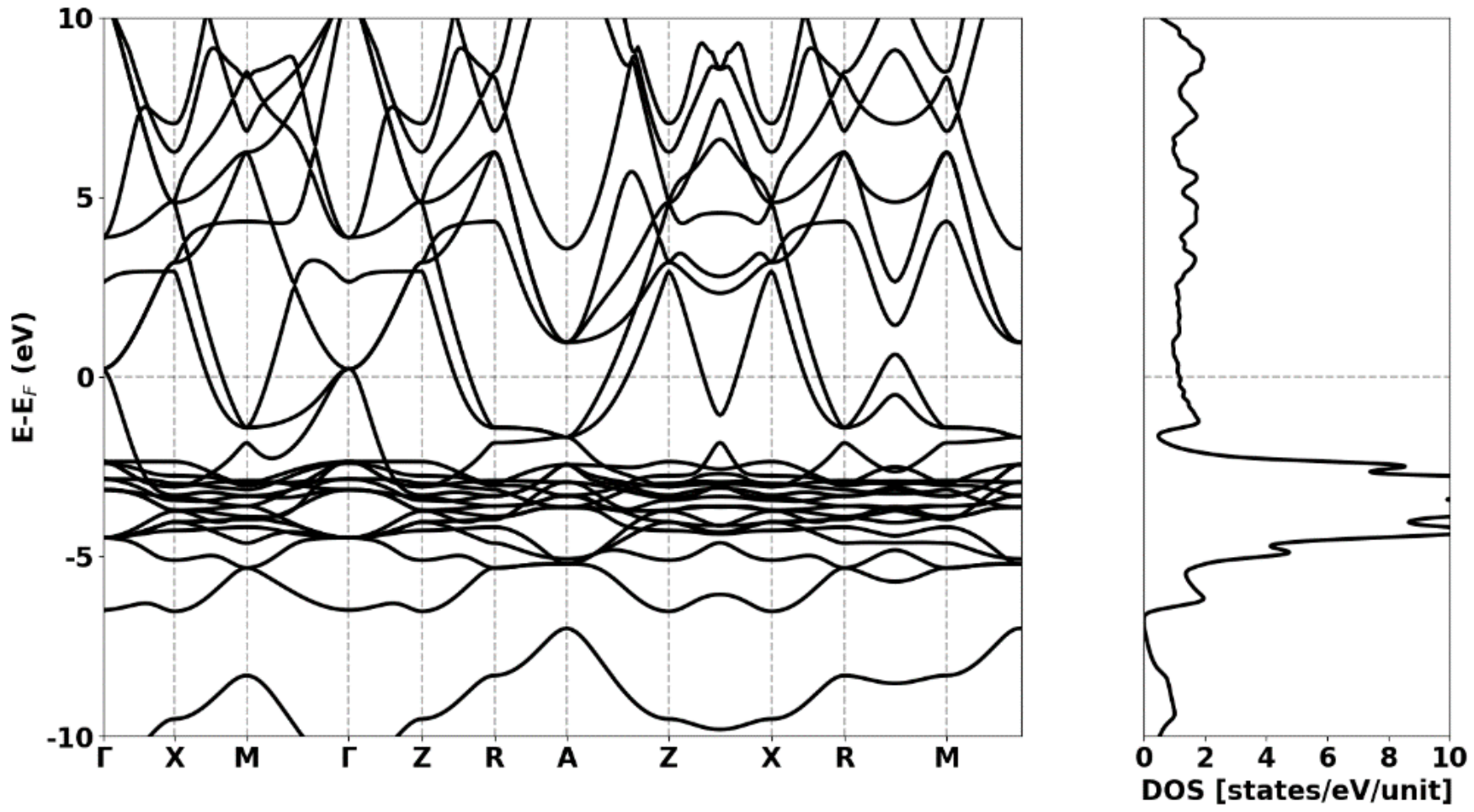

PBEsol 18x18x18

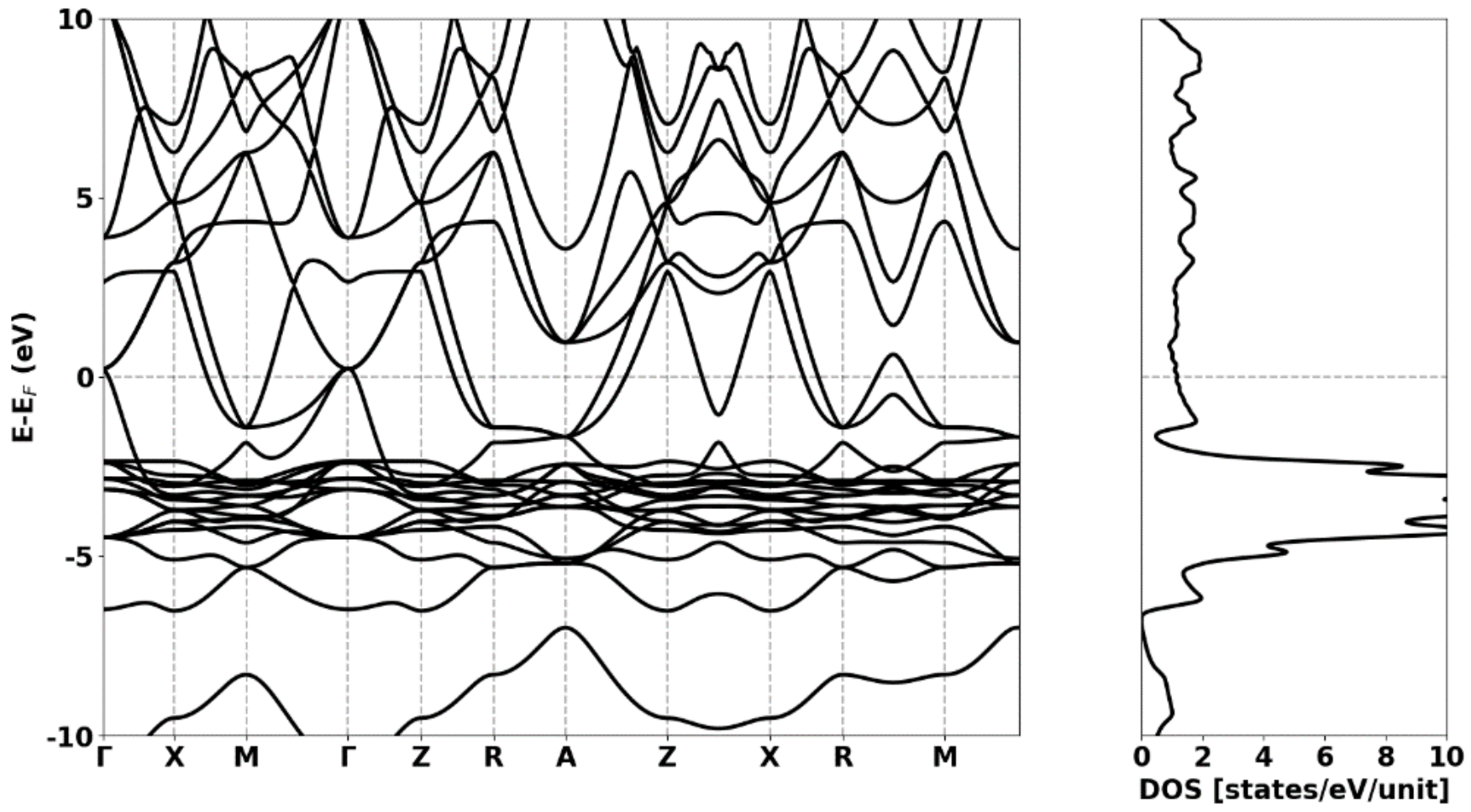

PBEsol 24x24x24

Not determined

S2. Phonon band structure computed with PBE or PBEsol functionals and four different increasing grid densities, as well as phonon density of states and Eliashberg function (whenever structure is dynamically stable)..

PBE 12x12x12

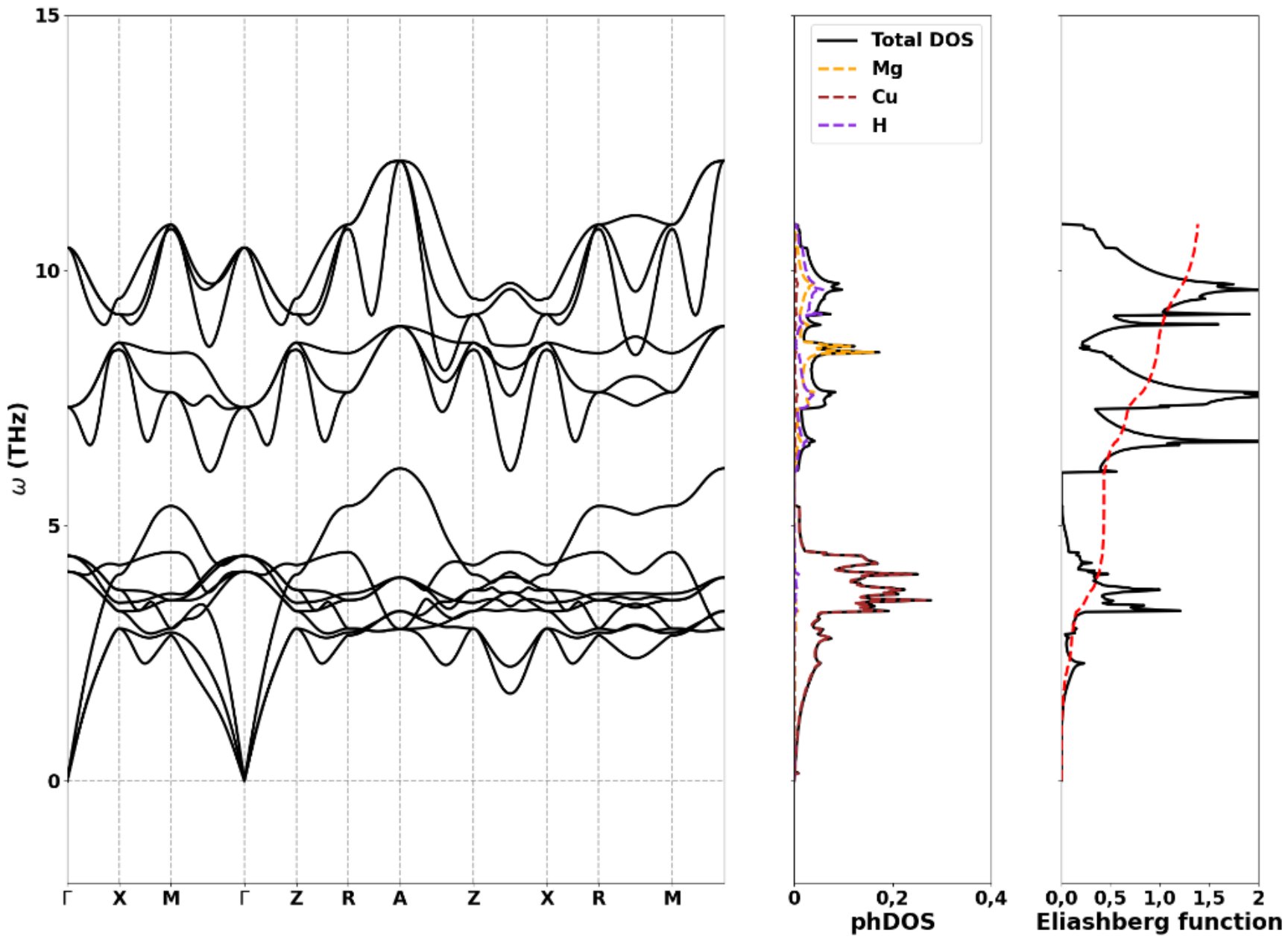

PBE 15x15x15

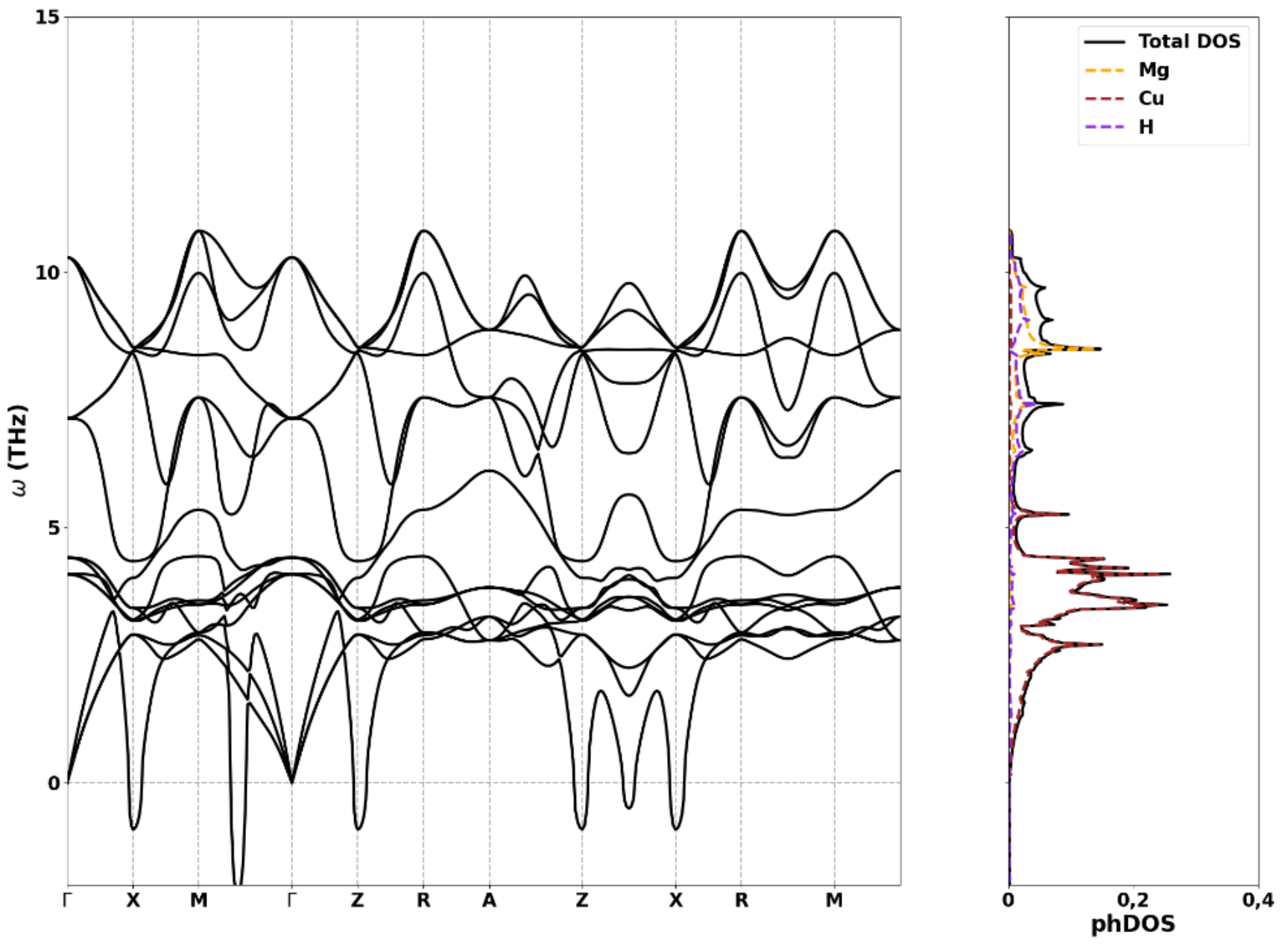

PBE 18x18x18

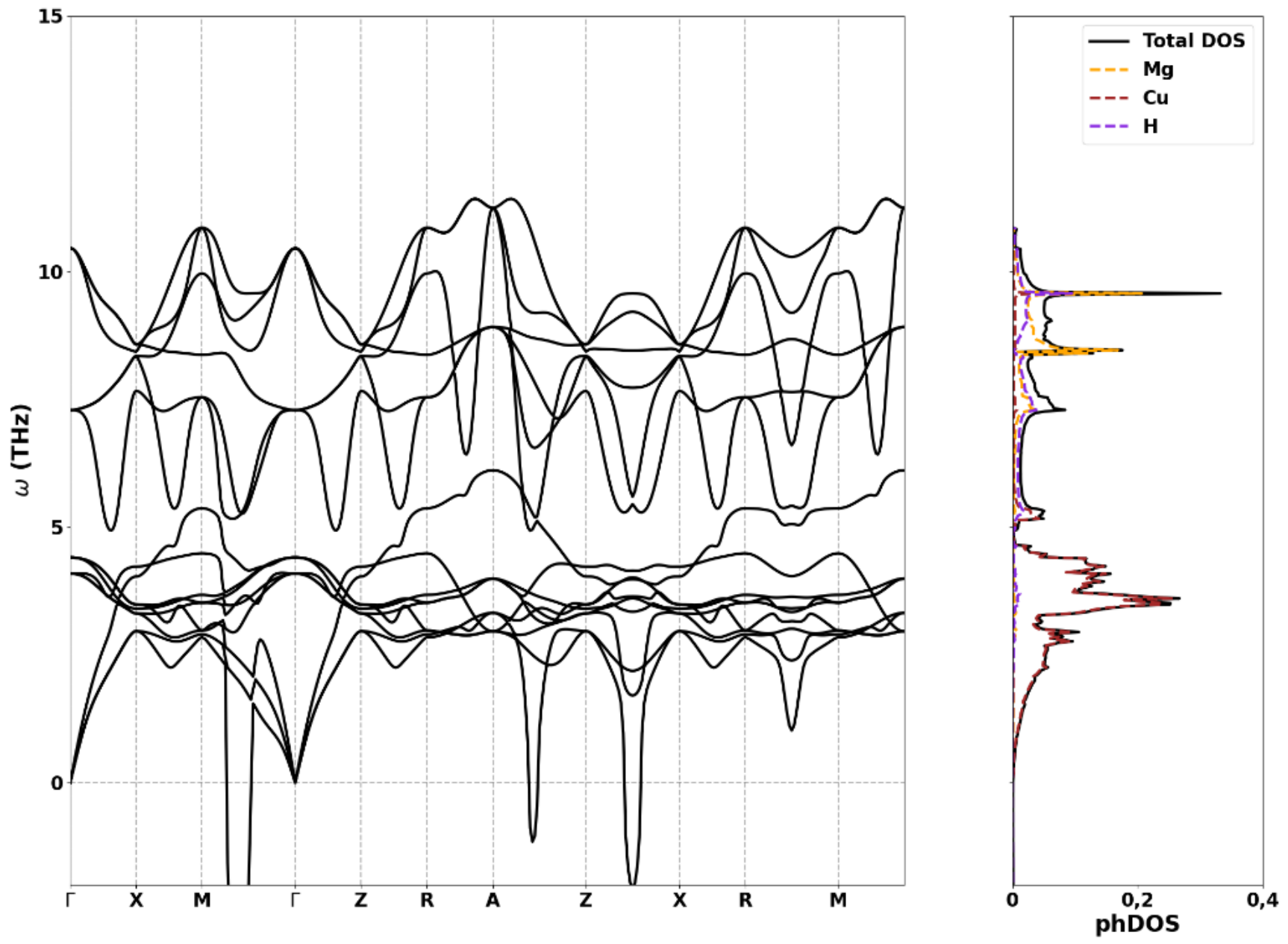

PBE 24x24x24

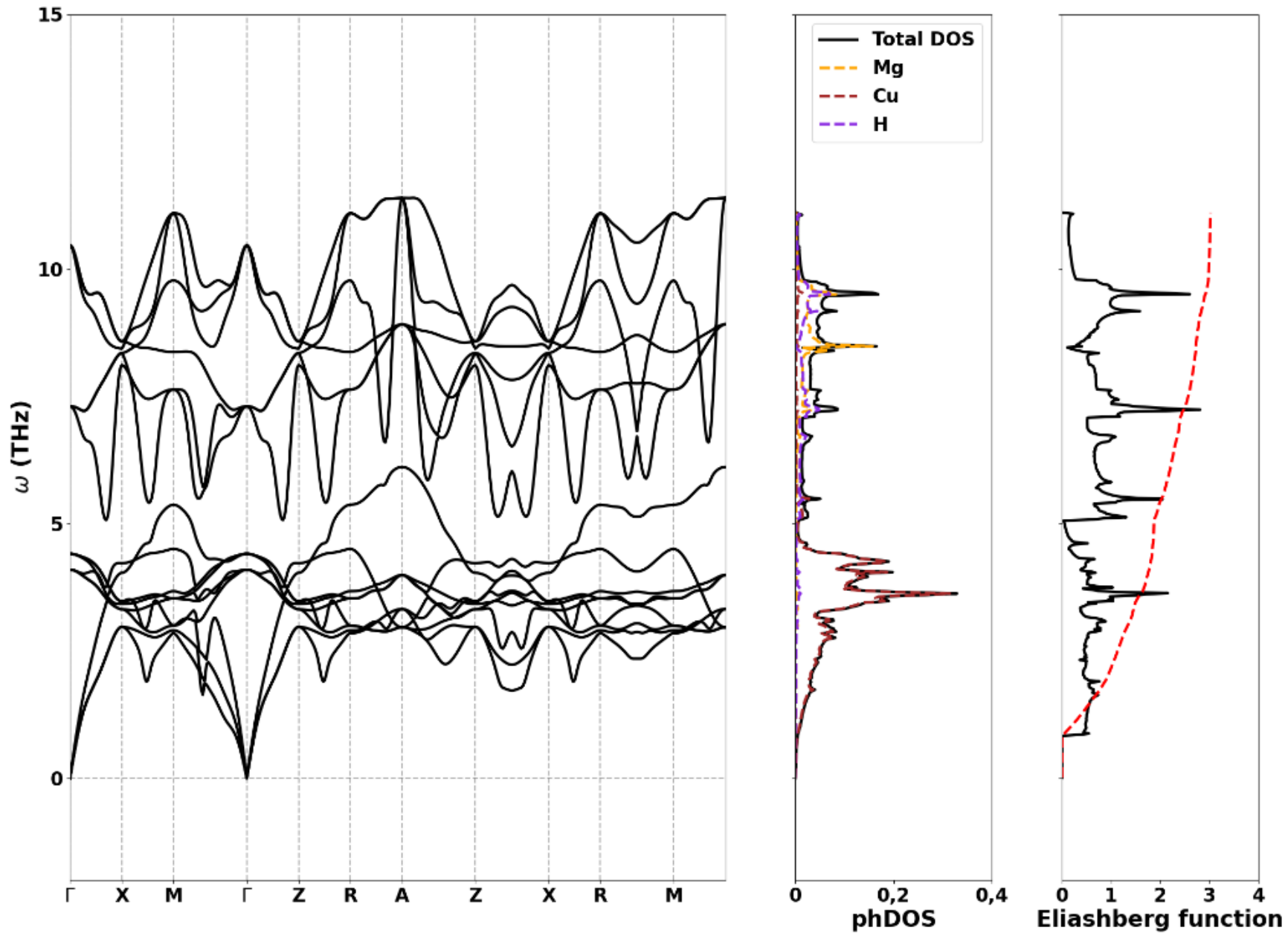

PBEsol 12x12x12

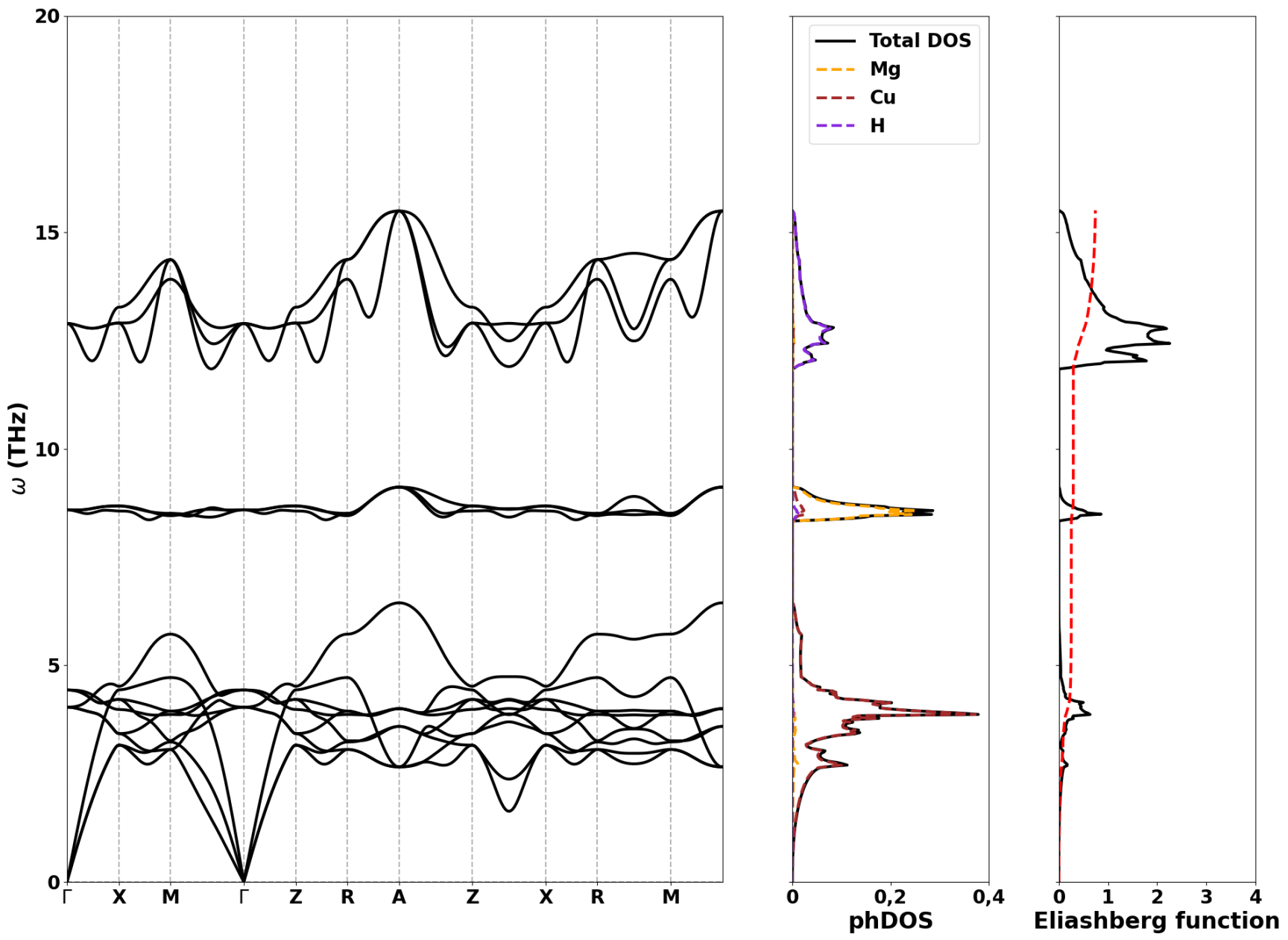

PBEsol 15x15x15

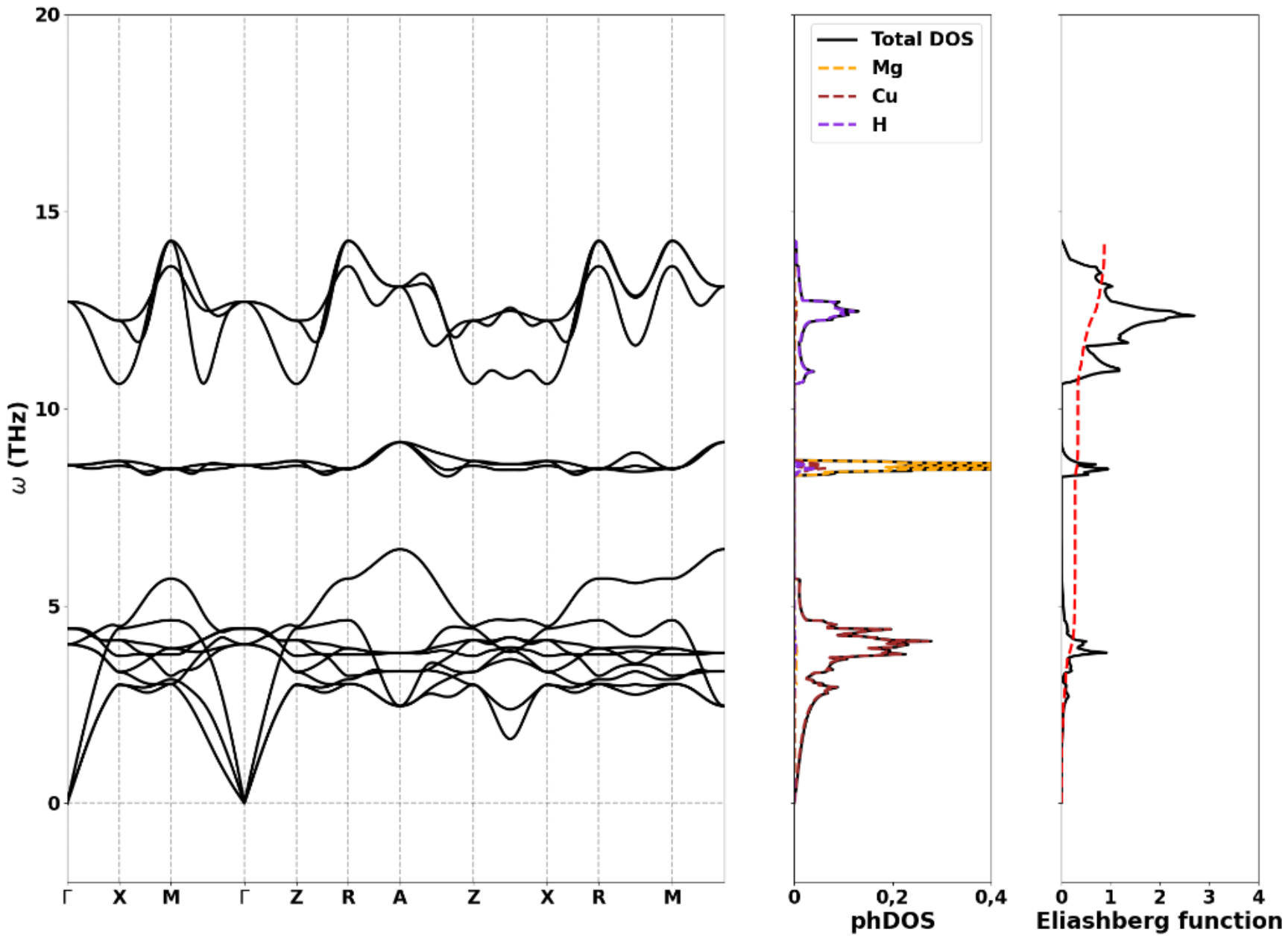

PBEsol 18x18x18

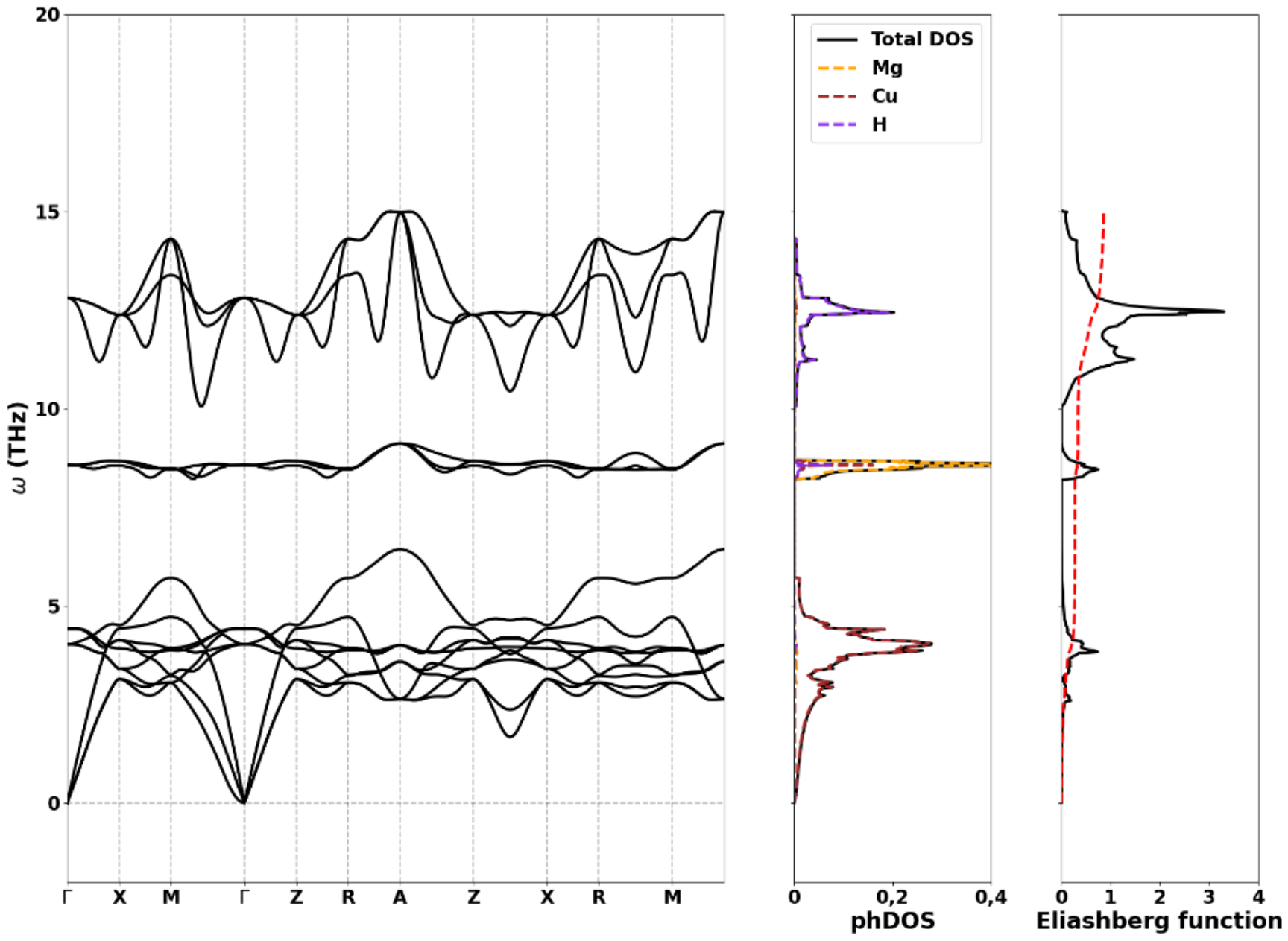

PBEsol 24x24x24

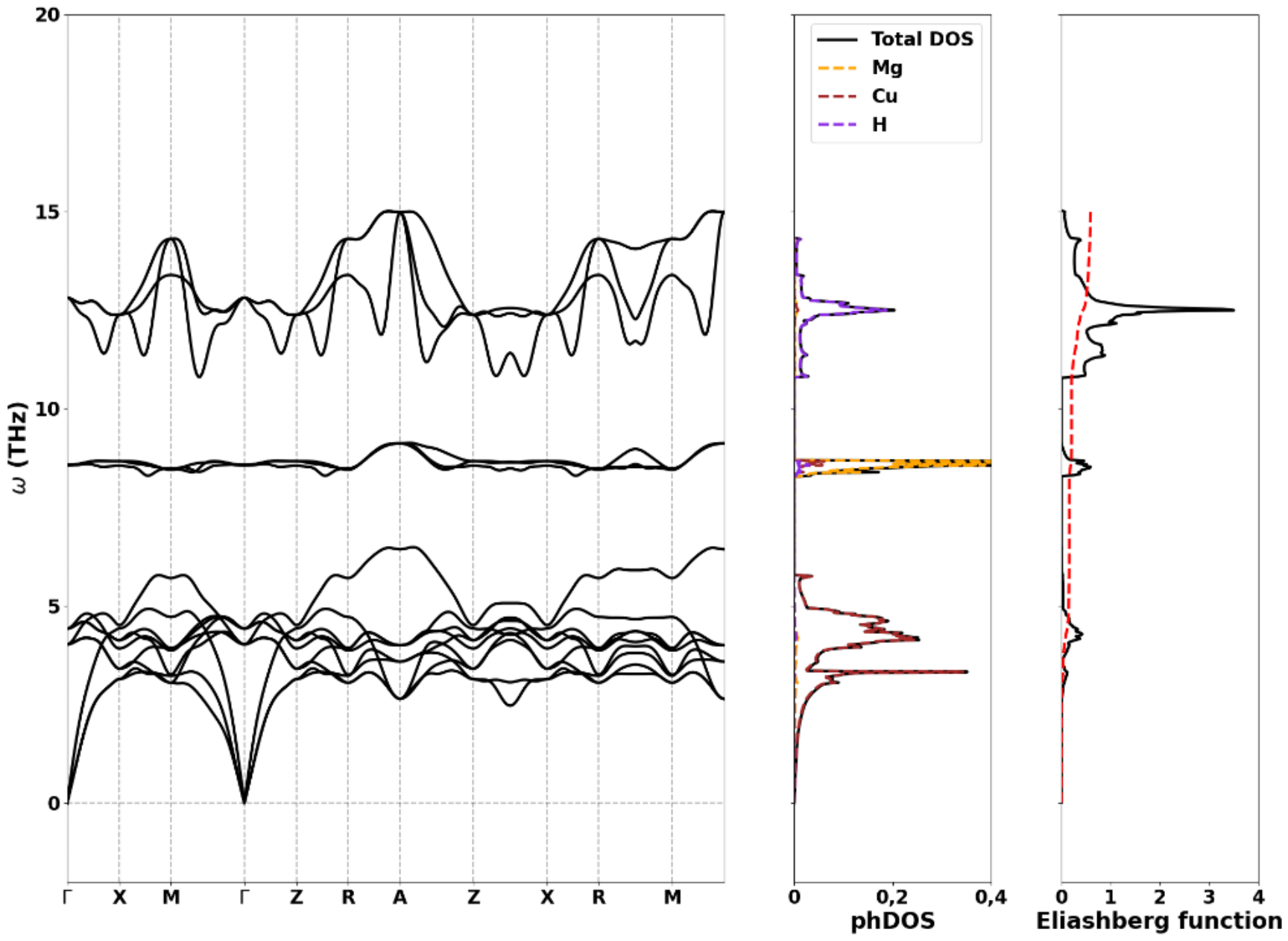

S3. The $T_C$ values in function of the Gaussian broadening computed with PBE or PBEsol functionals and four different increasing grid densities (whenever structure is dynamically stable).

PBE 12x12x12

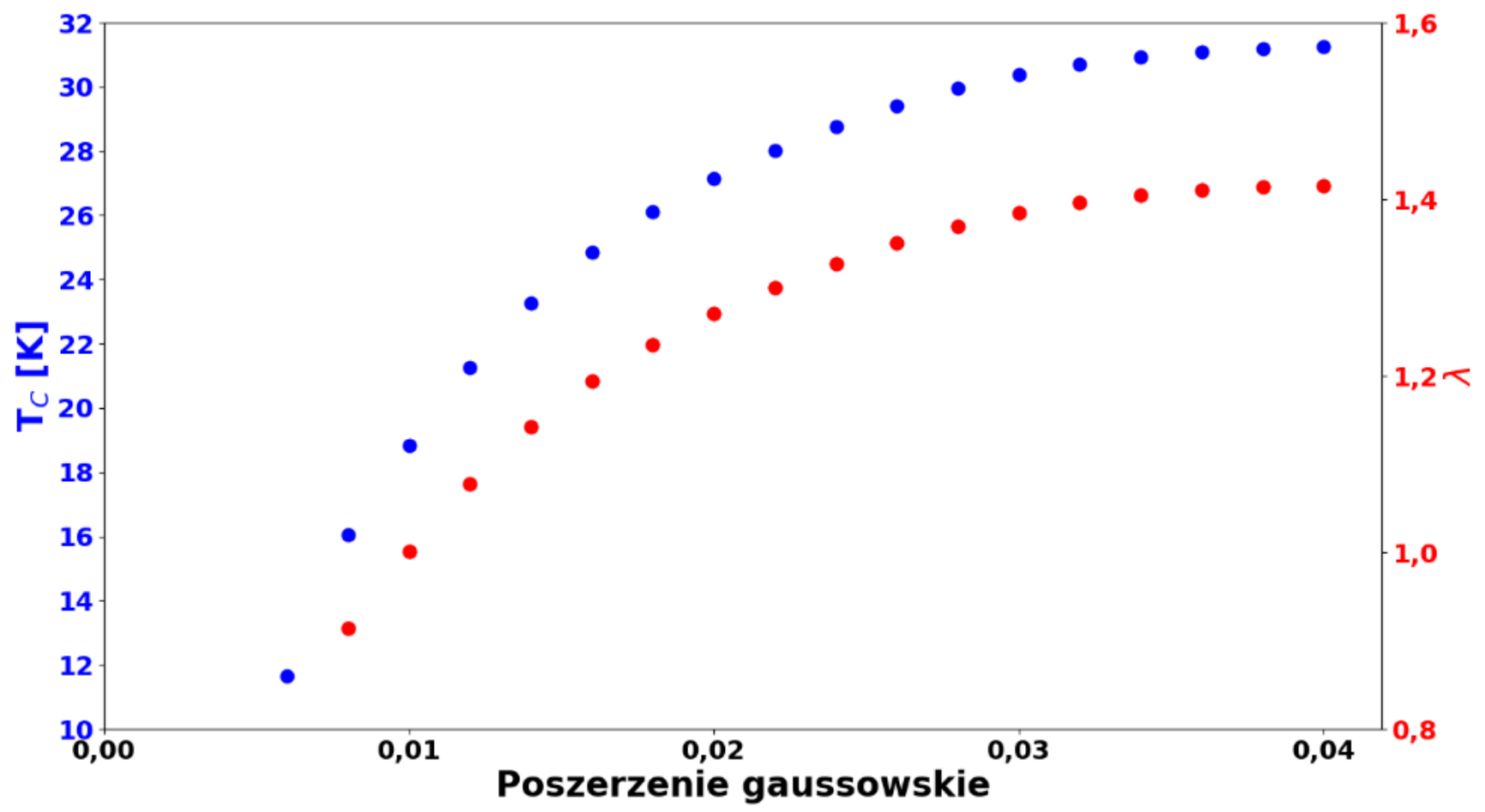

PBE 24x24x24

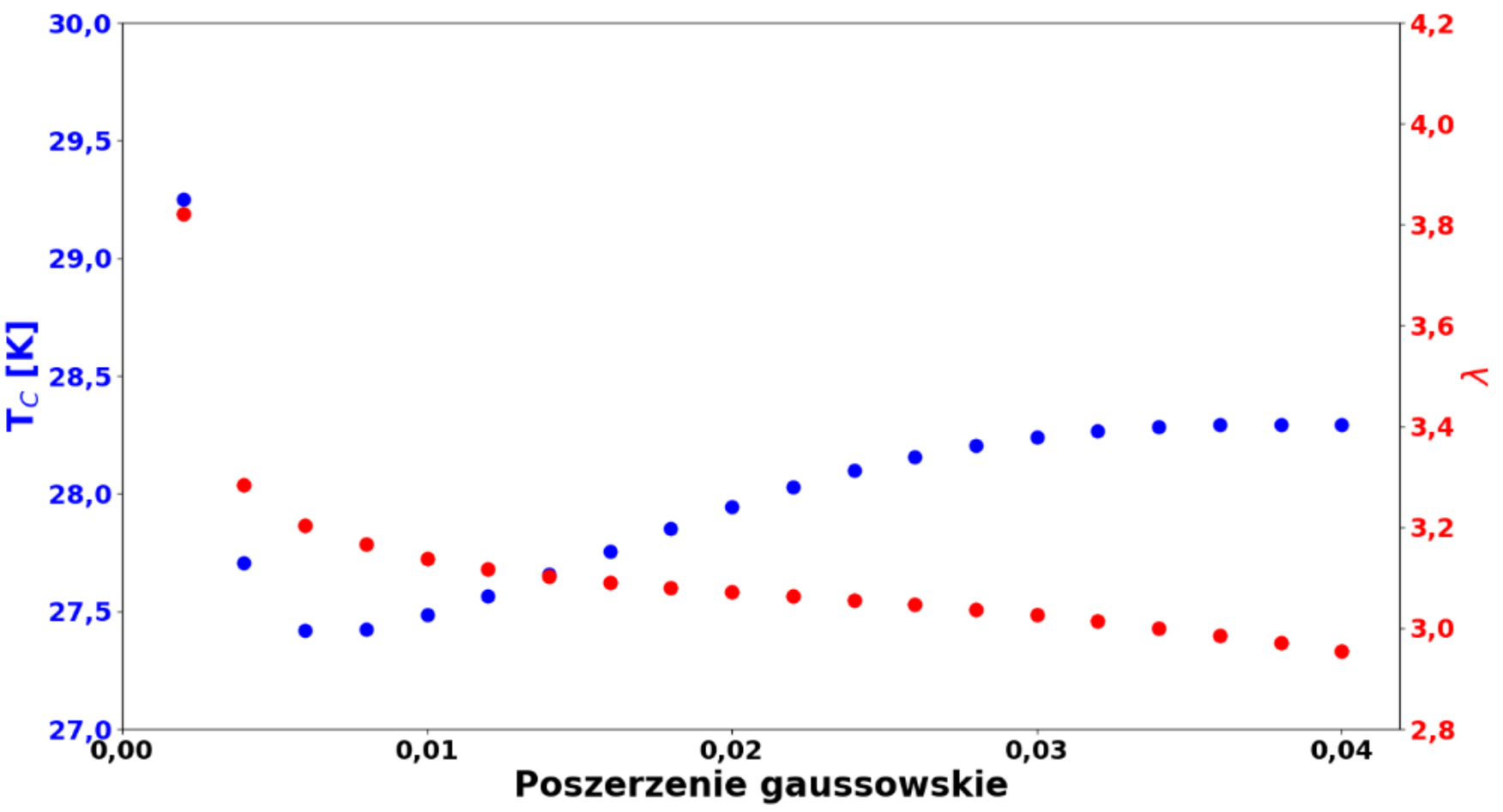

PBEsol 12x12x12

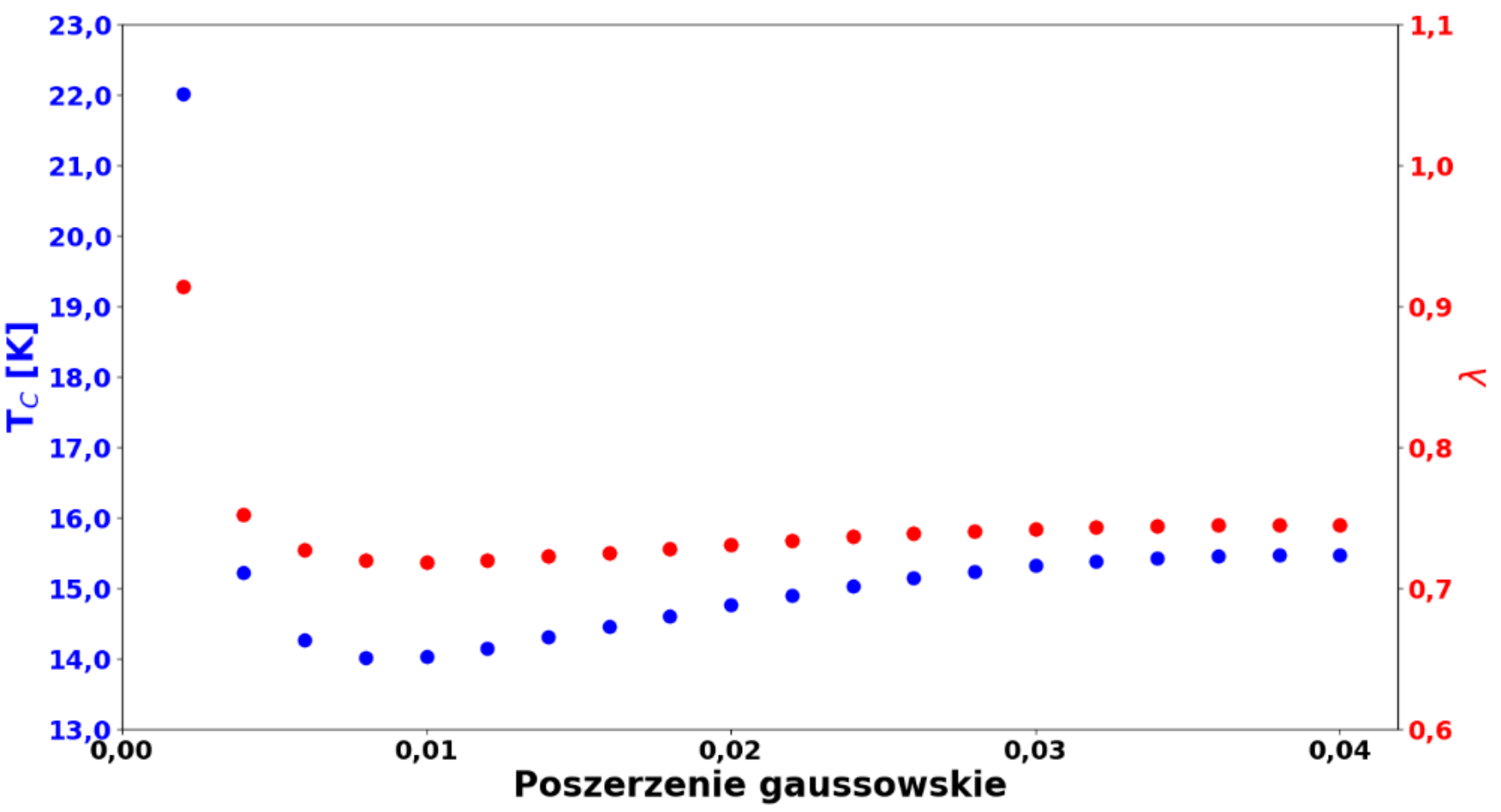

PBEsol 15x15x15

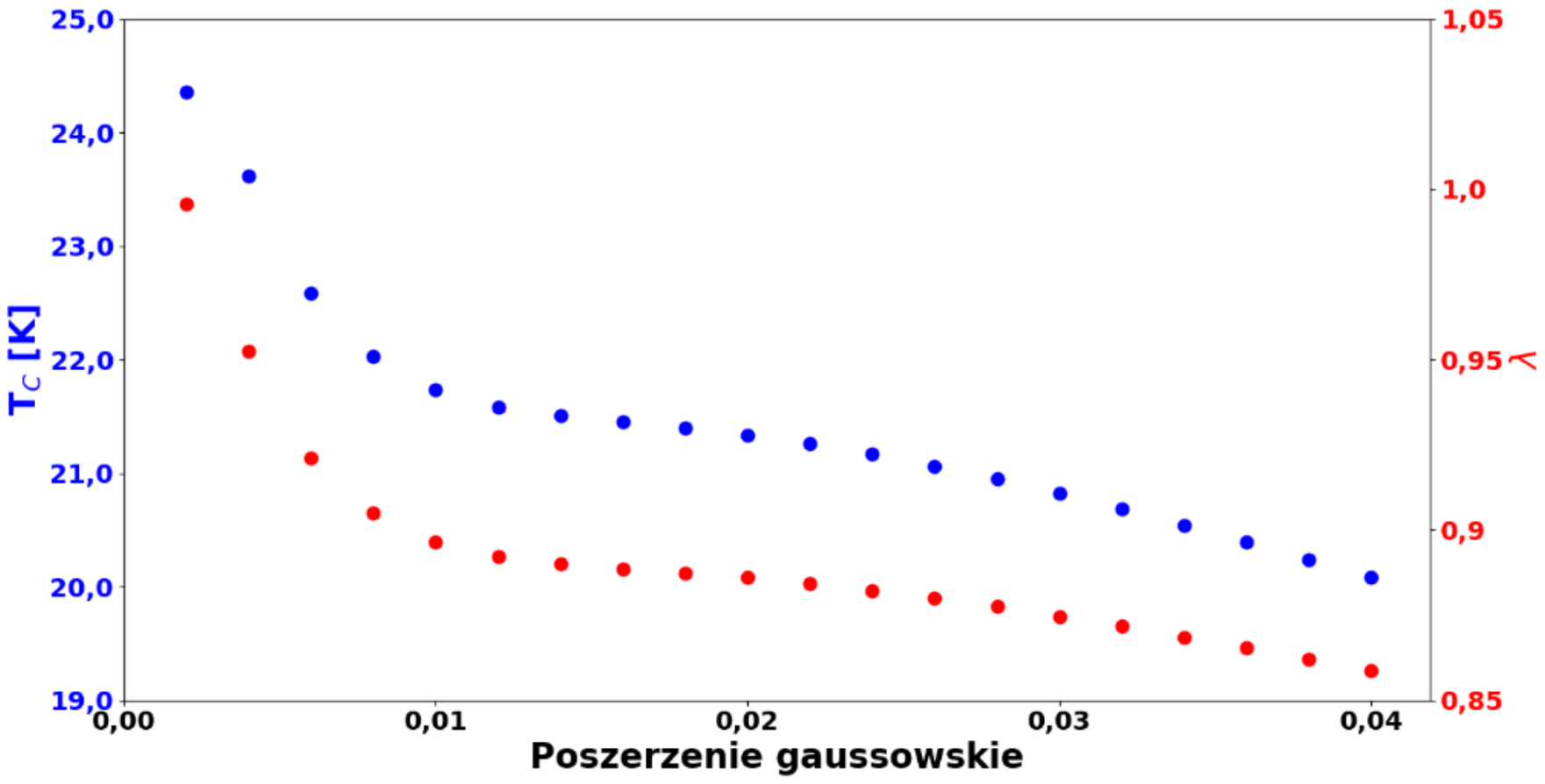

PBEsol 18x18x18

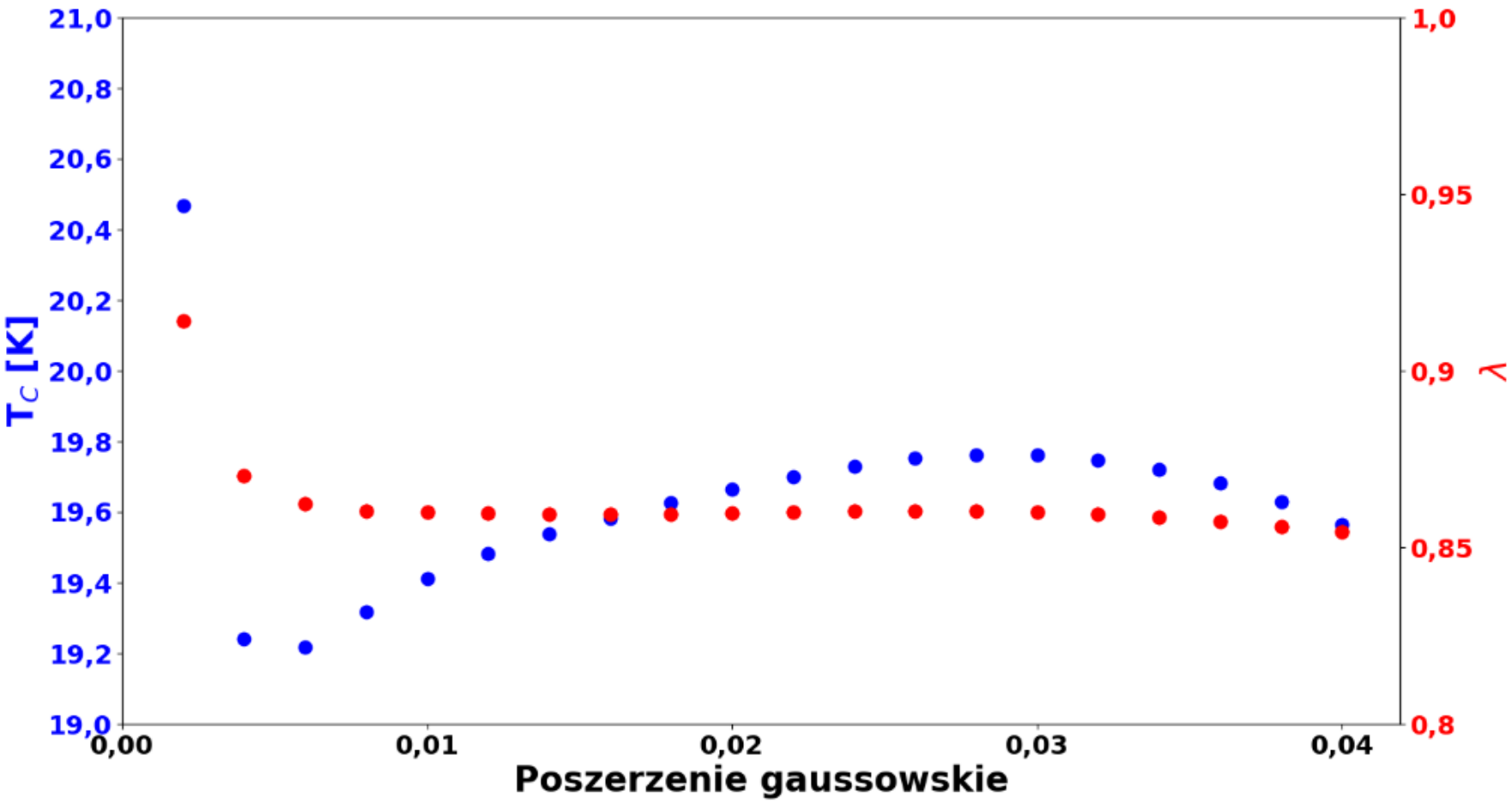

PBEsol 24x24x24

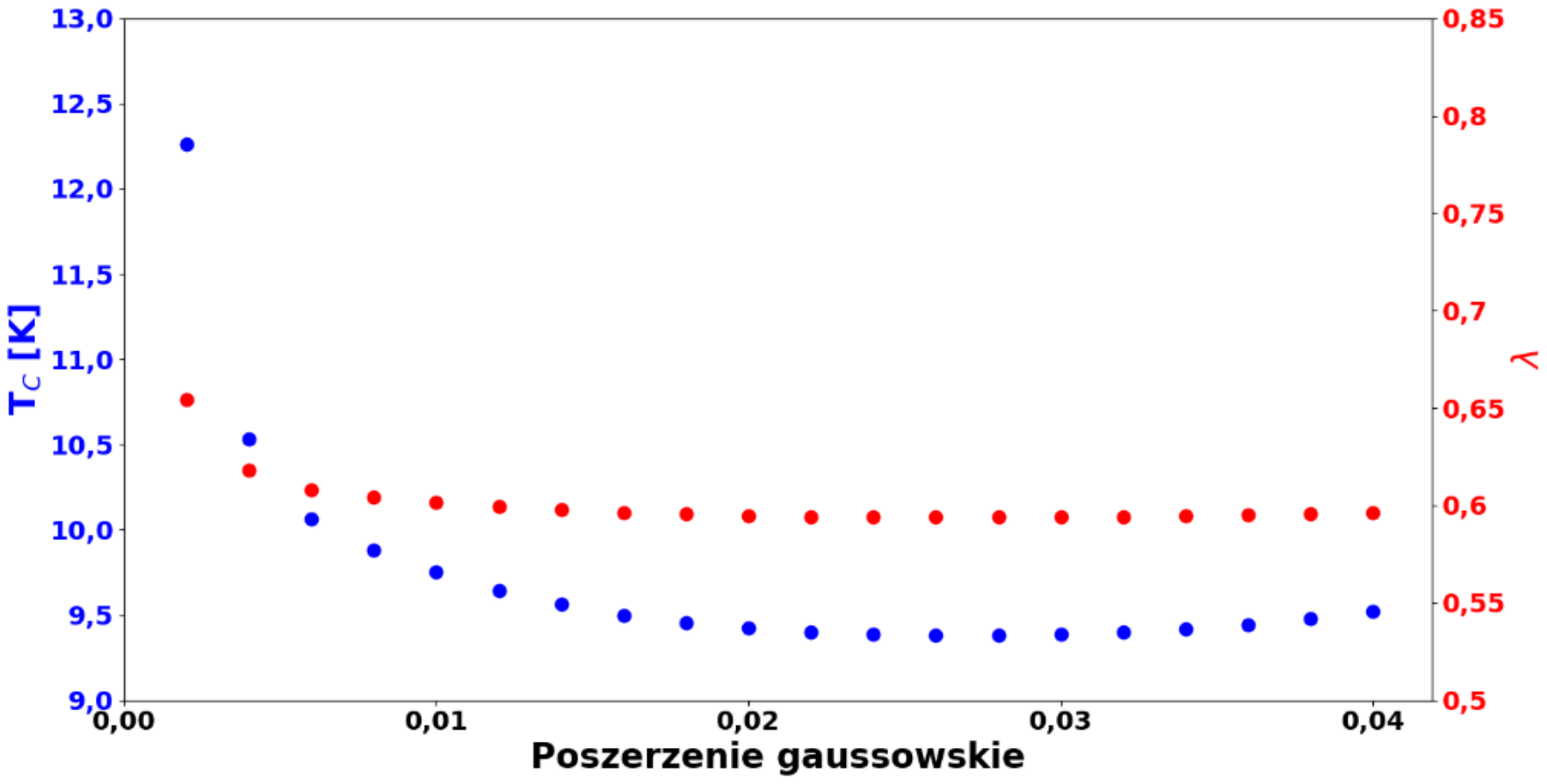